\documentclass[aps,prl,reprint]{revtex4-2}

\usepackage{amsmath,amsfonts,amssymb,graphicx,braket,bbold,mathtools,array}
\usepackage{float}
\usepackage{multirow}
\usepackage[]{units}
\usepackage{hyperref}
\usepackage{MnSymbol}
\hypersetup{
    colorlinks=true,
    linkcolor=blue,
    filecolor=blue,      
    urlcolor=blue,
    citecolor=blue
}

\usepackage[T1]{fontenc}

\newcommand*{\figref}[2][]{%
  \hyperref[{fig:#2}]{%
   \ref*{fig:#2}%
    \ifx\\#1\\%
    \else #1%
    \fi
  }%
}

\begin{document}
\title{Fingerprints of Qubit Noise in Transient Cavity Transmission}
\author{Philipp M. Mutter}
\email{philipp.mutter@uni-konstanz.de}
\author{Guido Burkard}
\email{guido.burkard@uni-konstanz.de}
\affiliation{Department of Physics, University of Konstanz, D-78457 Konstanz, Germany}

\begin{abstract}
Noise affects the coherence of qubits and thereby places a bound on the performance of quantum computers. We theoretically study a generic two-level system with fluctuating control parameters in a photonic cavity and find that basic features of the noise spectral density are imprinted in the transient transmission through the cavity. We obtain analytical expressions for generic noise and proceed to study the cases of quasistatic, white and $1/f^{\alpha}$ noise in more detail. Additionally, we propose a way of extracting the spectral density for arbitrary noise in a frequency band only bounded by the range of the qubit-cavity detuning and with an exponentially decaying error due to finite measurement times. Our results suggest that measurements of the time-dependent transmission probability represent a novel way of extracting noise characteristics.
\end{abstract}
\maketitle

A main threat to large-scale quantum computation is the decoherence of the qubit state~\cite{Zurek2003} which can be induced by a noisy environment~\cite{Huang2021}. Common noise sources include fluctuating electromagnetic fields that arise from noisy control parameters such as gate voltages as well as the interaction with the environment, e.g., 1/f noise due to two-state fluctuators in the material \cite{Paladino2014review}, the nuclear spin bath in semiconductor quantum dots~\cite{Kloeffel_review2013, Zhang_review2019, Burkard2021arXiv}, and magnetic flux noise, quasiparticles, and two-level fluctuators in superconducting circuits \cite{Clarke2008,Oliver2013,Krantz2019,deGraaf2020}. Noise even affects spin qubits in isotopically purified materials that suffer neither from the hyperfine interaction with the host nuclei nor directly from the gate voltages. This is because gate operations often rely on the spin-orbit interaction which couples the spin to the charge degree of freedom, thus introducing charge noise to the system~\cite{Huang2014, Benito2019b}. Relevant examples are hole spins in germanium where the spin-orbit interaction and hence possible spin-photon couplings are particularly strong~\cite{Hendrickx2020, Hendrickx2020b, Mutter2020cavitycontrol, Hendrickx2021four, Mutter2021natural, Jirovec2021, Mutter2021_ST_qubit, Jirovec2021_SOI_arXiv}.

With quantum error correction still a long way off, the suppression of noise is of the utmost importance in intermediate-scale quantum devices~\cite{Preskill2018}. As a first step, it is necessary to characterize the noise present in a given system~\cite{Forster2014, Shulman2014}. In this paper, we propose a novel way of extracting noise characteristics from the cavity transmission which we analyze using input-output theory \cite{Collett1984,Gardiner1985}. The transmission through a cavity in the steady-state has been used for the dispersive readout of the quantum state of superconducting qubits \cite{Blais2004,Schuster2005}, cavity photons \cite{Schuster2007}, and spin qubits \cite{Mi2018}, as well as for the determination of system parameters such as valley splittings~\cite{Burkard2016,Mi2017} and to detect signatures of the strong coupling regime of cavity quantum electrodynamics~\cite{Wallraff2004,Viennot2015, Stockklauser2017, Mi2018, Koski2020}. Here, we go beyond the steady-state case and demonstrate that fluctuations affecting the qubit leave a clear trace in the transient transmission. Our model makes few assumptions on the form of the two-level system and the noise and is thus generally applicable to a wide range of cavity-coupled qubit systems. In particular, it is well suited for describing the effects of fluctuating voltages on superconducting qubits and semiconductor charge qubits as well as fluctuating (effective) magnetic fields on spin qubits.

	\begin{figure}
		\includegraphics[scale=0.23]{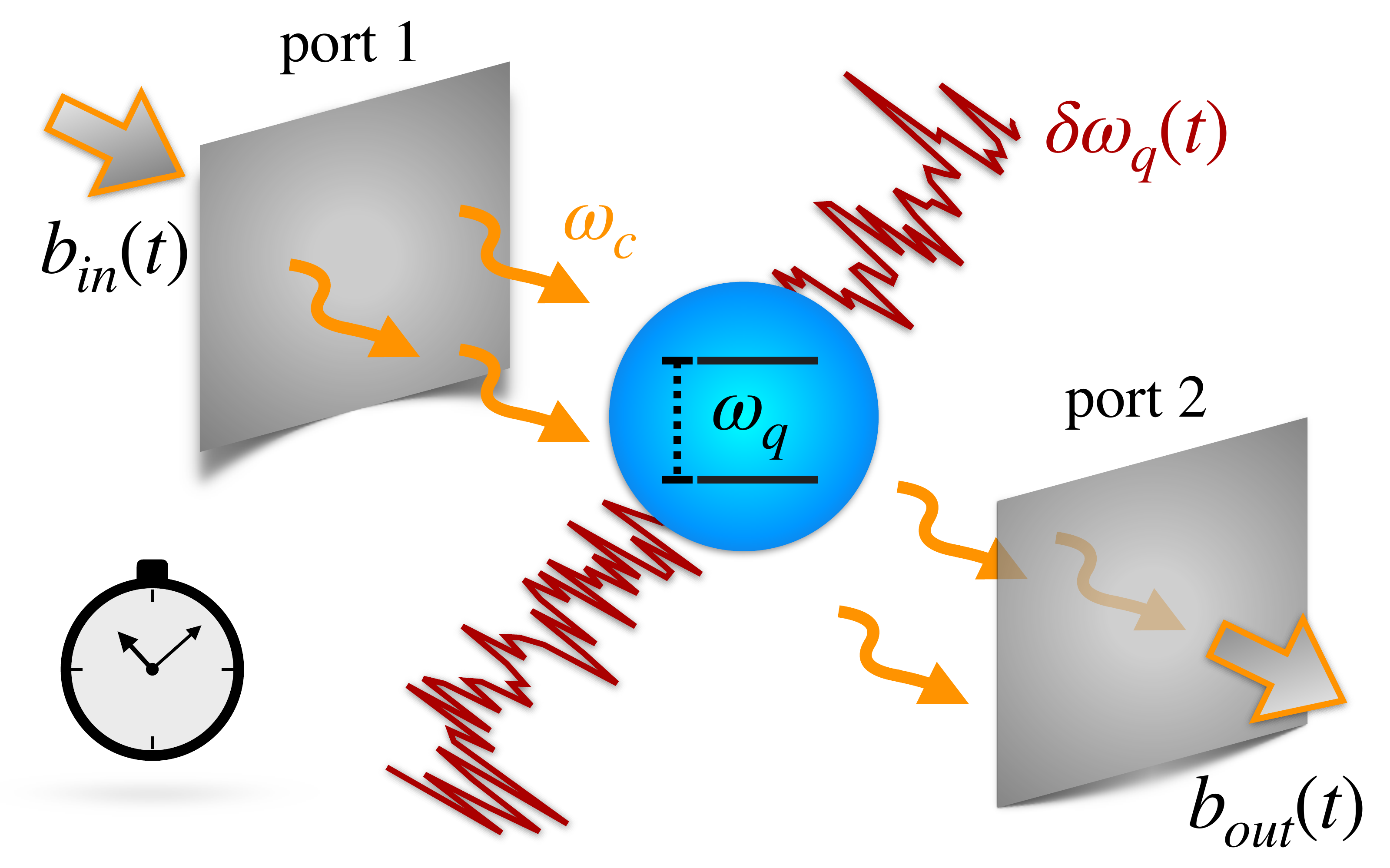}
		\caption{A two-level system (qubit, shown in blue) with energy splitting $\omega_q$ is affected by noise $\delta\omega_q$ (red) and is placed inside a single-mode electromagnetic cavity with frequency $\omega_c$. Partially transparent mirrors allow for the interaction of the qubit-cavity system with external modes. An input field $b_{\rm in}(t)$ enters at port 1, causing an output field  $b_{\rm out}(t)$ that leaves the cavity at port 2, thereby creating a time-dependent transmission through the system from left to right.}
		\label{fig:schematic_of_system}
	\end{figure}

\begin{figure}
		\includegraphics[scale=0.208]{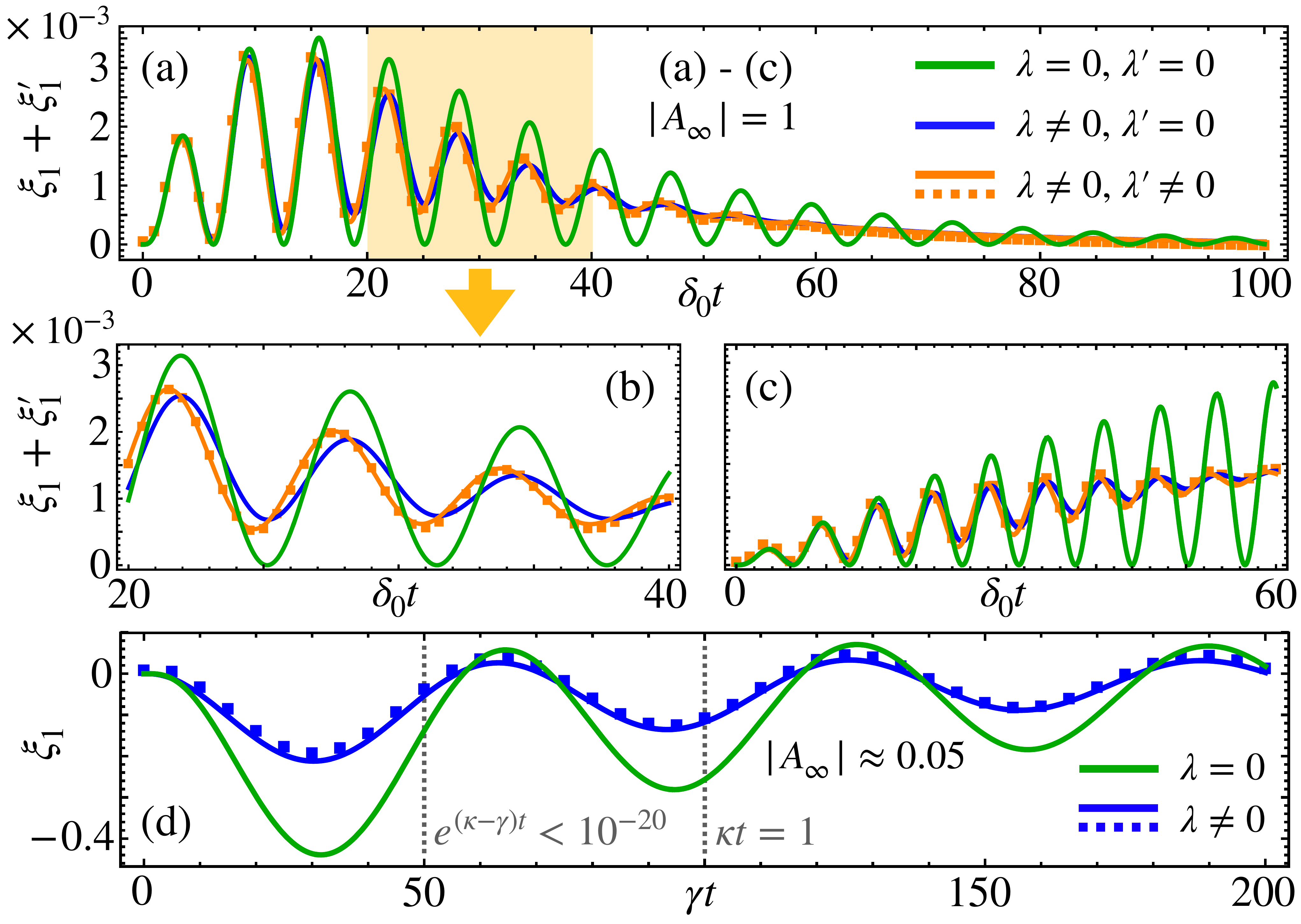}
		\caption{Change in the noise-averaged cavity transmission probability $\left\llangle \vert  A(t) \vert \right\rrangle=\llangle \vert\langle b_{\text{out}} (t) \rangle  / \langle b_{\text{in}} (t)\rangle\vert\rrangle$ in first order of the coupling $g$ to a noisy qubit as a function of time. We compare the cases of a noise-free (green) and noisy (blue/orange) qubit placed inside the cavity. Solid lines are drawn according to Eqs.~\eqref{eq:transmission_probability_terms} (blue) and~\eqref{eq:A_squared_additional_term} (orange). Squares are numerical results obtained by averaging over $10^3$ exact solutions of the full system of differential equations for $\langle a\rangle$ and $\langle\sigma_-\rangle$ (Eqs.~\eqref{eq:Langevin_equations}) and $\langle \sigma_z \rangle$ (see SM) assuming normally distributed quasistatic noise with zero mean and standard deviation $\delta X_{\text{rms}} = 0.05 \delta_0$. (a) The complete first order transient curve for the case $g = 0.1 \kappa = 0.1 \gamma  =0.01 \delta_0$ at $\Delta = 0$. The average transmission probability follows from $\xi_1$, $\xi_1'$, and $\vert A_\infty\vert$ via Eq.~\eqref{eq:transmission_probability}. We find excellent agreement between the analytical and numerical results, even in the presence of a fluctuating coupling constant as can be seen from the magnified section of the plot in panel (b). (c) The initial first order transient curve for the strong coupling case $g =  \kappa =  \gamma  = 0.01 \delta_0$ at $\Delta =0$. (d) The first order transient curve for $g = \kappa = 0.01 \gamma$ at $\delta_0  = \Delta = 0.1 \gamma$ with $\delta X_{\text{rms}} =  \gamma$. The remaining parameter values used are $\langle \sigma_z(0) \rangle =0$, $\lambda = 0.9$, $\lambda^{\prime} = -0.1$, $\omega_q/T = 1$ and $ \sqrt{\kappa} / \langle b_{\text{in}} \rangle =  1$. }
		\label{fig:transmission_comparison}
	\end{figure}

We study a single mode of an electromagnetic cavity with frequency $\omega_c$ interacting with an infinite number of external modes. A qubit affected by noise is placed inside the cavity and interacts with the cavity mode  (Fig.~\ref{fig:schematic_of_system}). In a frame co-rotating with the probe field at frequency $\omega_p$ and within the rotating wave approximation, the cavity-coupled two-level system is described by the Hamiltonian $H = [\omega_q + \delta \omega_q(t) - \omega_p ] \sigma_z/2 + \Delta a^{\dagger} a + g (a \sigma_+ + a^{\dagger} \sigma_-) $, where $\Delta = \omega_c - \omega_p$ is the cavity-probe detuning, $\sigma_z$ the Pauli $Z$ matrix, $a$ the photon annihilation operator and $\sigma_-$ a ladder operator acting on the qubit. The qubit energy separation is affected by noise, $\omega_q + \delta \omega_q(t)$, and the fluctuating component may be written to leading order as $\delta \omega_q(t) = \lambda \delta X(t)$, where $\delta X(t)$ is the dynamical noise which couples to the qubit control parameter $X$ with strength $\lambda = \partial_X \omega_q \vert_{\delta X = 0}$. The qubit-cavity coupling can also be affetced by noise, and we will take this into account further below. As is shown in the supplementary material (SM), the system is well described by the quantum Langevin equations for the expectation values of the operators $\sigma_-$ and $a$,
	\begin{align}
	\label{eq:Langevin_equations}
		\frac{\text{d} \langle \sigma_- \rangle}{\text{d}t} & = -i \left[  \omega_q + \delta \omega_q(t) - \omega_p  \right]  \langle \sigma_- \rangle - \frac{\gamma}{2}\langle \sigma_- \rangle   + i g \langle \sigma_z \rangle \langle a \rangle, \nonumber \\
		\frac{\text{d} \langle a \rangle}{\text{d}t} & = -i \Delta  \langle a \rangle - \frac{\kappa}{2} \langle a \rangle - i g \langle \sigma_- \rangle  + \sqrt{\kappa_1} \langle b_{\text{in}}(t) \rangle,
	\end{align}
where $\gamma$ is the total noise-independent qubit decoherence rate~\footnote{{As is shown in the SM, the qubit decoherence rate reads $\gamma = \coth(\omega_q/2T) \gamma_1 + 2 \gamma_{\varphi}$ with zero-temperature relaxation rate $\gamma_1$ and dephasing rate $\gamma_{\varphi}$. When solving the full system of five Langevin equations to obtain the numerical data points in Fig.~\ref{fig:transmission_comparison} we set $(\gamma_1, \gamma_{\varphi}) = (5  \tanh(\omega_q/2T), 2.5) 10^{-2} \delta_0$ in (a) and (b), $(\gamma_1, \gamma_{\varphi}) = (5 \tanh(\omega_q/2T), 2.5)10^{-3} \delta_0$ in (c), and $(\gamma_1, \gamma_{\varphi}) = (0.5  \tanh(\omega_q/2T), 0.25) \gamma$ in (d).}}, and $\kappa = \kappa_1 + \kappa_2$ is the total cavity loss rate given by the sum of the rates $\kappa_j$ at port $j \in \lbrace 1,2 \rbrace$. $\langle \sigma_z \rangle$ can depend on time but knowledge of its specific form is not required for computing the transmission within the perturbative approach presented below. The bath input field at port 1 is chosen to be a plane wave and hence a constant in the rotating frame, $\langle b_{\text{in}}(t) \rangle = \langle b_{\text{in}} \rangle$, while we assume no input field to be present at port 2.

By solving the system of coupled differential equations~\eqref{eq:Langevin_equations}, one may obtain the cavity transmission amplitude by employing input-output theory~\cite{Collett1984,Gardiner1985,Burkard2020},
	\begin{align}
	\label{eq:transmission_amplitude}
		A(t) = \frac{\langle b_{\text{out}} (t) \rangle  }{ \langle b_{\text{in}} (t)\rangle} =  - \frac{\sqrt{\kappa_2} \langle a (t) \rangle  }{ \langle b_{\text{in}}(t) \rangle}.
	\end{align}
An exact solution of the equations of motion cannot be obtained for generic noise, and one must resort to perturbation theory. Working in the regime where the qubit cavity coupling is small compared to the dominant energy scale, $ \vert g \vert  \ll \eta \equiv  \text{max} \lbrace\vert \delta_0 \vert, \vert \kappa - \gamma \vert \rbrace$ with the unperturbed qubit cavity detuning $\delta_0 = \omega_c - \omega_q$, we may solve the system of differential equations~\eqref{eq:Langevin_equations} to leading order in $g/\eta$,
	\begin{align}
	\begin{split}
	\label{eq:A_first_order_g}
	    A(t)   & = \frac{ \sqrt{\kappa_1 \kappa_2}}{i \Delta +  \kappa/2} \left(e^{-i \Delta t -\kappa t /2} - 1 \right)  \\
		& \quad -  \frac{\sqrt{\kappa_2}}{\langle b_{\text{in}} \rangle} e^{-i \Delta t - \kappa t/2}  \left[ \langle a(0) \rangle - i g   \langle \sigma_-(0) \rangle \mathcal{I} (t) \right], \\
		 \mathcal{I}(t) & = \int_0^t e^{i \delta_0 t^{\prime} + (\kappa - \gamma) t^{\prime}/2 -i \lambda \mathcal{X}(t^{\prime})}  d t^{\prime}.
	\end{split}
	\end{align}
Here, $\langle \sigma_-(0) \rangle$ and $\langle a(0) \rangle$ are initial conditions, and we introduce the noise integral $\mathcal{I}(t)$ containing the stochastic phase $\mathcal{X}(t) = \int_0^t \delta X(t^{\prime} ) \text{d} t^{\prime}$. There are a few remarks in order here: (i) Even though we treat $g$ perturbatively, this does not mean that our approach does not contain strong coupling cases with $\vert g \vert > \kappa, \gamma$. In such cases the approximation remains sound if the qubit cavity detuning is large enough, $\vert g \vert \ll \vert \delta_0 \vert$. (ii) To leading order in perturbation theory the long time solution is unchanged in the presence of noise, and only the transient transmission allows for a determination of noise characteristics. (iii) Even at this point we can see the role of the initial qubit state. For $\langle \sigma_-(0) \rangle$ to be non-vanishing, we need the qubit to be initialized in a coherent superposition of its energy eigenstates, and in the following we assume $\langle \sigma_-(0) \rangle = 1/2$. Moreover, we assume that $\langle a(0) \rangle = 0$, e.g., the cavity may initially be empty. (iv) There are two quantities in~\eqref{eq:Langevin_equations} that are affected by finite temperature effects: the qubit level population $\langle \sigma_z (t) \rangle$ and the decoherence rate $\gamma$.  The former appears in the expansion of $\langle a (t) \rangle$ only at higher orders in perturbation theory, and the latter is only altered in magnitude  at increased thermal energies, while the form of the Langevin equations is unchanged (see SM). As a result, Eq.~\eqref{eq:A_first_order_g} also describes the noisy transmission at finite temperature. Remarkably, temperature does not wash out the noise traces in the transient cavity transmission in magnitude. However, an increased $\gamma$ can lead to a quickly decaying noise signal, hence requiring small measurement times.  

Averaging over the noise is possible once we consider an observable quantity, such as the transmission probability $\vert A \vert ^2$ that will be investigated here. We remark that since the zeroth order term in~\eqref{eq:A_first_order_g} is not affected by noise one has $\llangle \vert  A \vert^2 \rrangle = \left\llangle \left\vert  A \right\vert \right\rrangle^2$ to first order in $g/\eta$, and hence the variance of $\vert A \vert$ vanishes. In general, the $k$th central moment of $\vert A \vert$ can become non-zero only at order $k$ or higher in $g/\eta$, implying $\llangle \vert  A \vert^k \rrangle = \left\llangle \left\vert  A \right\vert \right\rrangle^k + \mathcal{O}((g/\eta)^2)$ (a proof is given in the SM). In present-day two-level systems one may expect the decay rate $\gamma$ to be of the same order as the cavity loss rate and the qubit photon coupling constant, $\gamma \sim \kappa \sim g \sim$~MHz~\cite{Mi2018}. For the perturbative approach to be valid we then must consider the dispersive regime, $\vert \delta_0 \vert  \gg \vert g \vert$. Assuming symmetric mirrors $\kappa_1 = \kappa_2 = \kappa/2$ and choosing $\langle b_{\text{in}} \rangle$ to be real, we obtain up to leading order in the qubit-cavity coupling strength,
	\begin{align}
	\label{eq:transmission_probability}
		 \left\llangle \vert  A(t) \vert \right\rrangle = \vert A_{\infty} \vert \sqrt{\xi_0(t) + \xi_1(t)}, 
	\end{align}
Here, $\vert A_{\infty} \vert = \kappa/\sqrt{4\Delta^2 + \kappa^2}$ is the Lorentz shaped transmission through an empty cavity at long times $\kappa t \gg 1$, and we introduce the quantities	
	\begin{align}
	\label{eq:transmission_probability_terms}
		 \xi_0(t) & =    1 + e^{- \kappa t} - 2 e^{-\kappa t/2} \cos \Delta t , \nonumber \\
		 \xi_1(t) & =     \frac{  g  e^{-\kappa t/2} }{ \sqrt{2/\kappa}  \langle b_{\text{in}} \rangle} \bigg[ F(t) \text{Re} \left\llangle \mathcal{I}(t)  \right\rrangle  + G(t) \text{Im} \left\llangle \mathcal{I}(t)  \right\rrangle \bigg], \nonumber \\		
		 F(t) & = \frac{2\Delta}{ \kappa } \left( \cos \Delta t - e^{-\kappa t/2} \right) - \sin \Delta t, \nonumber \\
		 G(t) & = \cos \Delta t - e^{-\kappa t/2} + \frac{2\Delta}{ \kappa } \sin \Delta t.
	\end{align}
$\xi_0(t)$ describes the transient signal of an empty cavity, while $\xi_1(t)$ is a correction term due to the interaction with the noisy qubit. The latter features the averaged noise integral (ANI),
	\begin{align}
	\label{eq:definition_ANI}
		\left\llangle \mathcal{I} (t) \right\rrangle = \int_0^t e^{i \delta_0  t^{\prime} + (\kappa- \gamma) t^{\prime} /2} \left\llangle e^{-i \lambda \mathcal{X}(t^{\prime})} \right\rrangle  d t^{\prime},
	\end{align}
where $\llangle \dots \rrangle$ denotes the average over many measurements. In order to neglect the $g^2$ term when expanding the absolute value squared of the transmission amplitude~\eqref{eq:A_first_order_g} while still suppressing higher orders in the perturbation expansion, we require $\sqrt{\kappa} / \langle b_{\text{in}} \rangle \sim 1$. This restriction, however, is not severe as the amplitude $\langle b_{\text{in}} \rangle$  of the probe field may be tuned externally and independently of the remaining parameters. The expression in Eq.~\eqref{eq:transmission_probability} is one of the main analytical results of this paper. It describes the transmission amplitude for quite general systems and without any specifications of the noise $\delta  X(t)$ or the corresponding stochastic phase $\mathcal{X}(t)$, and it is shown for exemplary parameter settings in Fig.~\ref{fig:transmission_comparison}. By recording the noisy part of the transmission via comparison with the transmission through an empty cavity for two distinct detunings $\Delta_1$ and $\Delta_2$, one may extract the real and imaginary part of the ANI for any $\delta_0$ up to a desired maximum time $t_m$ by choosing $ 
\vert \Delta_2 - \Delta_1 \vert t_m  < \pi$  (see SM).

Up to this point the linearized noise has been treated exactly. In many realistic systems, however, noise affecting the qubit energy separation will also affect the qubit photon coupling constant $g$. Writing $g \rightarrow g + \delta g(t)$, where to leading order $\delta g(t) = \lambda^{\prime} \delta X(t)$ with $\lambda^{\prime} = \partial_X g \vert_{\delta X = 0}$, we may work to first order in perturbation theory in this additional noisy qubit photon coupling and integrate by parts to obtain an additional term in the averaged transmission (see SM),
	\begin{align}
	\label{eq:transmission_probability_squared_with_g_term}
		\llangle  \vert A(t) \vert \rrangle  = \vert A_{\infty} \vert \sqrt{\xi_0(t) + \xi_1(t) + \xi'_1 (t)},
	\end{align}
where $\xi_0(t)$ and $\xi_1(t)$ are as given in Eq.~\eqref{eq:transmission_probability_terms} and
	\begin{align}
	\label{eq:A_squared_additional_term}
	\begin{split}
		&\xi'_1(t) =\frac{\lambda^{\prime}}{\lambda}  \frac{ e^{-\kappa t/2 }}{ \sqrt{2/\kappa}  \langle b_{\text{in}} \rangle}  \mathcal{F}(t)
	\end{split}
	\end{align}
with
    \begin{align}
         \mathcal{F}(t)  = \; &  e^{ (\kappa - \gamma) t/2} \left \llangle e^{-i \lambda \mathcal{X}(t)} \right \rrangle  \left[ G(t) \cos \delta_0 t  - F(t) \sin \delta_0 t \right] \nonumber \\
		&  +  G(t)  \left[  \delta_0  \text{Im} \left\llangle \mathcal{I}(t) \right\rrangle   - \frac{\kappa - \gamma}{2} \text{Re} \left\llangle \mathcal{I}(t) \right\rrangle - 1 \right] \nonumber \\
		&  +  F(t) \left[  \delta_0  \text{Re} \left\llangle \mathcal{I}(t) \right\rrangle  + \frac{\kappa - \gamma}{2} \text{Im} \left\llangle \mathcal{I}(t) \right\rrangle  \right].
    \end{align}
While we must treat the noise in $g$ perturbatively, this is not a severe restriction since the relative fluctuations in $g$ are expected to be bounded by the coupling itself, $ \delta g =\lambda^{\prime} \delta X < g$. The effect of noise in the energy separation and the qubit photon coupling on the averaged transmission can be clearly seen in Fig.~\figref[(b)]{transmission_comparison}. At the double resonance $\Delta = \delta_0 = 0$ the term $\xi_1$ vanishes identically, and for weak coupling $\xi'_1$ is the dominant contribution.

The proposed model is general in the sense that it describes any two-level system with fluctuating energies and coupling constants to first order in the noise parameter. For instance, for charge qubits the eigenenergy separation is determined by the dot detuning $\epsilon$ and the tunnel coupling $t_c$, $\omega_q = \sqrt{\epsilon^2 + 4 t_c^2}$, while the qubit-photon coupling strength scales as $g \propto t_c/\sqrt{\epsilon^2 + 4 t_c^2}$ \cite{Burkard2020}. If the detuning has a fluctuating component $\delta \epsilon (t)$, one finds
	\begin{align}
	\label{eq:fluctuating_energy_charge_qubit}
		\lambda = \frac{\partial \omega_q}{\partial \epsilon} \Big\vert_{\delta \epsilon = 0} =  \frac{\epsilon}{\omega_q}, \quad \lambda^{\prime} = \frac{\partial g}{\partial \epsilon} \Big\vert_{\delta \epsilon = 0} = - \frac{g \epsilon}{\omega_q^2}.
	\end{align}
For biased dots, $\epsilon \neq 0 $, a charge qubit affected by fluctuating gate voltages represents a prime example of a system described by our model. Note that the ratio $\vert \lambda^{\prime}/\lambda \vert = \vert g \vert /\omega_q $ can be reduced by increasing the dot detuning, and at large $\epsilon$ Eq.~\eqref{eq:transmission_probability} is a good approximation. Additionally, higher-order noise terms $\delta X^{n \geqslant 2}$ in the expansion of $\delta \omega_q$ couple with strength $\lambda^{(n)} = \partial^n \omega_q/\partial \epsilon^n \vert_{\delta \epsilon =0} \sim 1/\epsilon^{n+1}$ and are hence suppressed in this regime. For lower dot detunings, Eq.~\eqref{eq:A_squared_additional_term} can be taken into account to provide accurate corrections for $\lambda^{\prime} \delta X \lesssim g$.

Having obtained an expression for the measurable average transmission probability for generic longitudinal qubit noise affecting the energy separation and the coupling constant, we proceed to study the averaged phase (AP) $\llangle e^{-i \lambda \mathcal{X}(t)} \rrangle$ and the ANI $\llangle \mathcal{I}(t) \rrangle$ in more detail. The stochastic phase $\mathcal{X}(t)$ is defined as the time integral over the noise $\delta X(t)$ which is assumed to have zero mean in the remainder of this paper. When the autocorrelations $\llangle \delta X(0) \delta X(\tau) \rrangle$ decay on time scales $\tau_c$ which are small compared to the time of integration, the random phase is a sum of many independent random variables. In this situation the central limit theorem guarantees that the probability distribution of $\mathcal{X}(t)$ is Gaussian, and we may write for the AP~\cite{Chirolli2008, Bergli2009}
	\begin{align}
	\label{eq:noise_in_Gaussian_approximation}
		\left\llangle e^{-i \lambda \mathcal{X}(t)} \right\rrangle  =  \exp \left[ - \frac{\lambda^2}{2\pi} \int_{0}^{\infty} \frac{\sin^2 (\omega t/2)}{(\omega/2)^2} S(\omega) d \omega \right],
	\end{align}
where $S(\omega)$ is the noise spectral density, given by the Fourier transform of the noise auto correlator. Since the lower integration bound is zero, the Gaussian approximation is not expected to hold at measurement times $t$ that are of the same order as the noise correlation time $\tau_c$. Instead, the condition $\tau_c \ll t$ must be met for Eq.~\eqref{eq:noise_in_Gaussian_approximation} to accurately describe the stochastic phase in the transient cavity transmission. On the other hand, $\delta X(t)$ itself may be the sum of many uncorrelated microscopic modes. In this case  $\mathcal{X}(t)$ will follow Gaussian statistics regardless of the integration time $t$. If none of the above conditions are met, one must go beyond the Gaussian approximation~\cite{Makhlin2004}.

There are two prominent special cases for which the ANI in the Gaussian approximation may be explored further analytically, quasistatic noise and white noise. We first consider the case of quasistatic noise~\footnote{{In this case one could solve the Langevin equations~\eqref{eq:Langevin_equations} exactly by matrix diagonalization. However, the expression for the transmission probability becomes cumbersome and averaging analytically becomes impossible.}}. Assuming the total integration time $t$ to be smaller than the time scale on which the quasistatic noise changes, the ANI becomes a Gaussian and may be evaluated,
	\begin{align}
	\label{eq:quasistatic_noise_integral}
		 \left\llangle \mathcal{I}(t) \right\rrangle_{\rm qs} = \sqrt{\frac{\pi}{2}}  \frac{e^{Y^2}}{\lambda \delta X_{\text{rms}}}  \left[ \text{erf} \left(  Y \right) + \text{erf} \left(  \frac{ \lambda \delta X_{\text{rms}} t}{ \sqrt{2} }-   Y  \right) \right], 
	\end{align}
where $Y = [ i \delta_0 + (\kappa- \gamma)/2]/ \sqrt{2} \lambda \delta X_{\text{rms}}$, $\delta X_{\text{rms}} = \sqrt{\llangle \delta X^2 \rrangle}$ is the root mean square of the noise, and $\text{erf}$ denotes the error function. We now turn to the case of white noise, where $S = S_0$ is constant. The exponential of the AP becomes linear in time, yielding the exact expression for the ANI,
	\begin{align}
	\label{eq:white_noise_integral}
		\left \llangle \mathcal{I}(t) \right \rrangle_w  = \frac{e^{ i \delta_0 t + (\kappa - \gamma)t/2 - \lambda^2 S_0 t/2 } - 1}{i \delta_0 + (\kappa - \gamma)/2   -  \lambda^2 S_0/2 }.
	\end{align}
For $S_0=0$, Eq.~\eqref{eq:white_noise_integral} is the ANI in the noise-free case~\footnote{{Since the AP is bounded by unity due to its exponential form, the image of the real part of the noise-free ANI provides a superset of the image of the real part of the noisy ANI irrespective of the noise details, $\text{Re} \llangle \mathcal{I}_{\text{noise-free}} (\mathbb{R} ) \rrangle \supseteq \text{Re} \llangle \mathcal{I} (\mathbb{R} ) \rrangle$. An analogous relation holds for the imaginary part of the ANI.}}. As white noise describes a Markovian process, it only renormalizes the qubit decoherence rate $\gamma$ stemming from the Lindblad formalism, $\gamma \rightarrow \gamma + \lambda^2 S_0$. 

Next, we investigate generic noise. In real measurements, data cannot be acquired over an infinitely broad frequency band. This can be taken into account by introducing an ultraviolet (UV) cutoff in the integral appearing in Eq.~\eqref{eq:noise_in_Gaussian_approximation}, i.e., by shifting the upper integration bound from infinity to the UV cutoff frequency $\omega_{\text{uv}}$. For sufficiently small measurement times where $\omega_{\text{uv}} t \ll 1$, the sine function in~\eqref{eq:noise_in_Gaussian_approximation} may be expanded around zero. Taking into account the leading term in the expansion, the exponential in the AP becomes quadratic in time and the ANI may be evaluated,
	\begin{align}
	\label{eq:small_times_noise_integral}
		&\left \llangle \mathcal{I}(t) \right \rrangle = \frac{\pi}{\sqrt{2 P}} \frac{e^{Z^2}}{\lambda}  \left[ \text{erf} \left(  Z \right) + \text{erf} \left(   \sqrt{\frac{P}{2 \pi}} \lambda t -   Z  \right) \right],
	\end{align}
where $  Z = \sqrt{ \pi/ 2P} [i\delta_0 + (\kappa - \gamma)/2]/ \lambda  $ and $P = \int S(\omega) \text{d} \omega$ is the noise power in the positive frequency band $[0,\omega_{\text{uv}}]$ under consideration. As noise in quantum computation must be considered over a large bandwidth corresponding to gate operation times, the noise power $P$ provides a practical figure of merit for the comparison of quantum information platforms~\cite{Kranz2020}. Realistically, the condition $\omega_{\text{uv}} t \ll 1$ can be fulfilled for power spectral densities dominated by low frequencies such as $1/f^{\alpha}$ noise~\cite{Petersson2010}, which is ubiquitous in solid-state systems~\cite{Dutta1981review, Paladino2014review,Connors2019}. In these cases an additional infrared cutoff $\omega_{\text{ir}}$ is needed to regularize the power integral~\cite{Ithier2005,Russ2015}, justified, e.g., by finite data acquisition times~\cite{Nakamura2002}. For instance, for $S = C/\omega$ one has $P = C \ln (\omega_{\text{uv}}/\omega_{\text{ir}})$.

Finally, we consider arbitrary detunings $\delta_0$. Truncation of the perturbation expansion is valid in the regime $g \ll \vert  \kappa - \gamma \vert $, which is often realized through $\gamma \gg \kappa \sim g$ for long-lived coherence in solid-state qubits. Suppose that the ANI has been characterized for at least three values of the noise coupling strength $\lambda$ which is controllable by external parameters [e.g., via the detuning or tunnel coupling in charge qubits, cf. Eq.~\eqref{eq:fluctuating_energy_charge_qubit}]. One may then consider the second derivative of the ANI,
	\begin{align}
	\label{eq:second_derivative_ANI}
	\begin{split}
		& \frac{d^2 \llangle \mathcal{I}(t, \delta_0) \rrangle }{d \lambda^2} \bigg\vert_{\lambda = 0} = e^{i \delta_0 t + (\kappa - \gamma)t/2} \zeta(t) \\
		& \quad + \frac{16}{\pi (\kappa - \gamma + 2i \delta_0) } \int_0^{\infty} \frac{S(\omega)}{(\kappa - \gamma + 2i \delta_0)^2 + 4 \omega^2 }  d \omega ,
	\end{split}
	\end{align}
where $\zeta(t)$ is a function that is upper bounded by a quadratic scaling in $t$ (SM). Hence, the relative error caused by neglecting the first term is upper bounded by the scaling $\vert \kappa - \gamma \vert^2 t^2 e^{(\kappa - \gamma)t/2} $, and for measurement times $t$ with $  \kappa t \sim 1 \ll (\gamma - \kappa)t$ it can be safely neglected, while the effect of the noise on the transmission is still visible [Fig.~\figref[(d)]{transmission_comparison}]. By employing a partial fraction decomposition and using the symmetry of $S(\omega) = S(-\omega)$, the non-vanishing part of Eq.~\eqref{eq:second_derivative_ANI} may be rewritten as (see SM for details) 
	\begin{align}
	\label{eq:S_via_convolution}
	\begin{split}
		&\frac{d^2 \llangle \mathcal{I}( \delta_0) \rrangle }{d \lambda^2} \bigg\vert_{\lambda = 0} = \frac{16}{ (\kappa - \gamma + 2i \delta_0)^2} \mathcal{C}(\delta_0),
	\end{split}
	\end{align}
where $\mathcal{C}(\delta_0) = (S \star K)(\delta_0)$ denotes the convolution of the spectral density $S$ with the kernel $K(\delta_0) = (\kappa - \gamma + 2i \delta_0)^{-1}$. After Fourier transforming the kernel analytically, we may apply the convolution theorem to obtain the spectral density for arbitrary noise,
	\begin{align}
	\label{eq:S_via_convoltuion_theorem}
		&S(\delta_0)  = - 4  \int_0^{\infty} \tilde{\mathcal{C}}(\tau)  \cos (\delta_0 \tau) e^{(\gamma - \kappa) \tau/2 }d \tau,
	\end{align}
where $\tilde{\mathcal{C}}(\tau)$ denotes the Fourier transform of the measurable convolution. Hence, the spectral density can be obtained by extracting $\mathcal{C} (\delta_0)$ and its Fourier transform from the average of the transient cavity transmission and evaluating~\eqref{eq:S_via_convoltuion_theorem}.

In conclusion, we investigate the effect of noise in two-level systems on the cavity transmission and derive an analytical expression for the averaged transmission amplitude which depends on the type of noise present. To leading order in perturbation theory the noise signature is only visible in the transient but not in the long time transmission, depends on a true quantum mechanical initial qubit state, and is not limited by temperature. We investigate relevant types of noise and find that the transient cavity transmission carries information on characteristic features of the associated spectral density. This includes the root mean square of quasistatic noise, the amplitude of white noise and the power $P$ of noise types with low enough UV cutoff frequencies such as $1/f^{\alpha}$ noise. Finally, we propose a novel way of extracting the spectral density for arbitrary noise in a frequency band bounded only by the range of the qubit-cavity detunings and with an exponentially decaying error due to finite measurement times. Our results allow for the determination of noise characteristics in a wide variety of systems including cavity-coupled superconducting, charge and spin qubits.

\section{Acknowledgments}
This research is supported by the German Research Foundation [Deutsche Forschungsgemeinschaft (DFG)] under Project No. 450396347.

\bibliography{charge_noise_cQED}

\end{document}